\documentstyle[12pt,axodraw]{article}
\def\neu {\tilde\chi_1^0}
\def\l {\ell}
\def\slep {\tilde{\ell}}

\def\msl {m_{\slep}}
\def\mise {E \!\!\!\!/ ~~}

\def\beq{\begin{equation}}
\def\eeq{\end{equation}}

\newcommand{\lsim}
{{\;\raise0.3ex\hbox{$<$\kern-0.75em\raise-1.1ex\hbox{$\sim$}}\;}}
\newcommand{\gsim}
{{\;\raise0.3ex\hbox{$>$\kern-0.75em\raise-1.1ex\hbox{$\sim$}}\;}}

\begin{document}
%====================================================================%

\begin{flushright}
\texttt{hep-ph/0509161}\\
\texttt{TIFR/TH/05-35}\\[3ex]
\end{flushright}

\begin{center}
{\Large\bf  Event shape discrimination of supersymmetry from large extra dimensions at a linear collider} \\[10mm]
{\bf Partha Konar$^*$}
and {\bf Probir Roy}  \\[2ex]
{\em Department of Theoretical Physics, Tata Institute of Fundamental Research,\\ 
Homi Bhabha Road, Mumbai - 400 005, India.} \\
E-mail: {\sf konar@theory.tifr.res.in, probir@theory.tifr.res.in} \\ 
\end{center}
\vskip.5in

%====================================================================%
\begin{abstract}

The production of a charged lepton ($\l = e, \mu$) pair with a large
missing energy at a linear collider is discussed as a means of 
distinguishing the minimal supersymmetry (MSSM) scenario from that 
with large extra dimensions (ADD) for parameter ranges where the 
total cross sections are comparable for both. Analyses in terms of 
event shape variables, specifically sphericity and thrust, are shown 
to enable a clear discrimination in this regard.
\end{abstract}

\vskip 1 cm
\noindent
\texttt{{\bf \sc PACS Nos:} 04.50.+h, 11.10.Kk, 11.25.Mj, 12.60.Jv, 13.66.Hk, 14.80.Ly} \\

\noindent
\texttt{{\bf \sc Key Words}: Linear Collider, Beyond Standard Model, Supersymmetry Phenomeno\-logy, Large Extra Dimensions, Event Shape Variable.}\\

\vfill

{$^*$Address after October 1, 2005 : Institut f\"ur Theoretische Physik, Universit\"at Karlsruhe, D--76128 Karlsruhe, Germany}

\newpage

%====================================================================%
\noindent
{\underline {\em Introduction}}:~
A general expectation in high energy physics today is that of physics beyond the standard model (BSM) emerging at TeV energies. Supersymmetry (SUSY) \cite{Drees:2004jm} and extra dimensions \cite{Perez-Lorenzana:2005iv} are two alternative possibilities in this direction that are the most exciting. They both address the naturalness/gauge hierarchy problem, arising from quantum corrections to the Higgs parameters, via the introduction of new physics at the TeV scale. Moreover, their attractive phenomenological features, in particular their promise of new states a bit beyond the current experimental lower mass bounds, put them in the limelight among scenarios of BSM physics to be explored by search strategies presently being designed. The latter, in fact, constitute the major motivation for constructing the next generation of colliders. If either SUSY or an extra dimensional scenario should manifest itself at sub-TeV to TeV energies, its signals ought to show up at the upcoming Large Hadron Collider (LHC) at CERN. It is widely accepted, nonetheless, that the precise nature of the BSM physics responsible for such signals may not always be easily gleaned from analyses of the corresponding data on account of the complexity of the hadronic environment in any LHC process. Indeed, in order to unambiguously identify the nature and detailed properties of any such new physics, a high energy and high-luminosity $e^+ e^-$ machine \cite{LC} -- such as the proposed International Linear Collider (ILC) or the Compact Linear Collider (CLIC) -- will be very useful.

We consider the signal comprising unlike-sign dielectrons/dimuons, produced in a linear collider together with a very high amount of missing energy, seeking to distinguish between SUSY and and the Arkani-Hamed--Dimopoulos--Dvali (ADD) model \cite{ADD} of large extra dimensions\footnote{Within the extra dimensional paradigm, there are other scenarios such as warped (Randall-Sundrum) or universal extra dimensions, which we do not address here.}$^,$\footnote{Another process where the two scenarios have been compared is $e^+ e^- \to \gamma_H \mise\!\!\!\!$, where $\gamma_H$ is a hard photon. The reactions for the ADD and SUSY scenarios in standard notation are $e^+ e^- \to G_n \gamma_H$ and  $e^+ e^- \to \tilde{G} \tilde{G} \gamma_H$ \cite{Gopalakrishna:2001iv} or  $e^+ e^- \to \neu \neu \gamma_H$ \cite{Asakawa:2002ij} respectively. The energy spectrum of the hard photon together with the scaling of the cross section with CM energy and moment distributions of the transverse energy squared have been used for discrimination purposes. However, since there is only one observable particle in the final state, no event shape analysis is possible here.}. Such a process has already been considered  \cite{Bhattacharyya:2005vm,Battaglia:2005ma} in the context of the Universal Extra Dimension (UED) scenario \cite{Appelquist:2000nn}. The mechanism for this reaction is somewhat similar in SUSY and UED : a two-body production of heavy entities, each of which then has a dominant two-body decay. But the spins of the primarily produced entities are different in the two cases, leading to \cite{Battaglia:2005ma} distinguishable angular distributions and asymmetries. There are also differences in the lepton energy spectrum. We find, however that these quantities are not very sensitive to a SUSY vs ADD discrimination. First of all, the difference in these between the two scenarios is more quantitative, being in detailed shape aspects, rather than being something qualitative; systematic uncertainties would tend to wash out such quantitative differences. Secondly (and more importantly), these quantities are quite ISR-sensitive so that ISR-corrections significantly reduce the sensitivity to such a discrimination.

Let us give an illustration to highlight the last point. The famous box-shaped lepton energy spectrum in the SUSY case has been found (as shown in Fig. 5 of \cite{Battaglia:2005ma}) to be squeezed in energy, looking more like a peak, after ISR corrections. When we compare this corrected spectrum with the peaked one for the ADD case, there does not seem a whole lot of difference. Similarly,  the angular distributions are flat in either case for the bulk of the measurable range in the cosine of the angle between the two leptons. We do not include these plots here since that will detract from our central point which is the following. Distributions in event shape variables, such as sphericity and thrust, are known to be ISR-stable and are yet found to be sensitive to such a discrimination. They are ${\it qualitatively}$ different between SUSY and ADD, having a peak in sphericity or break in thrust for the former and monotonic fall or rise for the latter. This is owing to differences in the mechanisms leading to the $\l^+ \l^- \mise$ final state in the two cases. Of course, slepton pair-production for SUSY will have a distinct threshold in $\sqrt{s}$ unlike the generation of the corresponding ADD final state, the cross-section for which increases smoothly with $\sqrt{s}$. So a careful scan of the CM energy for a threshold will also help discriminate between the two. However, that will require a more detailed step-by-step analysis. It will be useful to have a discriminant just with the first set of data at a particular $\sqrt{s}$ (above the slepton pair production threshold) and this is what we provide.

We work within the minimal weak-scale R-parity conserving supersymmetric standard model (MSSM) which predicts the pair-production of charged sleptons \cite{slep_pair}, once the requisite energy threshold is reached, in an $e^+ e^-$ collider. Each produced slepton would perforce decay into a charged lepton and the lightest supersymmetric particle (LSP). The latter is normally taken to be the lightest neutralino $\neu$ which, being stable and interacting only weakly, escapes unobserved through the detector -- carrying a considerable amount of missing energy. In contrast, the ADD model has $d$ extra dimensions compactified on a $d$-torus. Together with time and the three spatial dimensions of our world, these constitute the bulk spacetime. The radius\footnote{For simplicity, we take the same radius of compactificaion for each of the $d$ dimensions.} $R_c$ of compactificaion of the extra dimensions could be as large as a quarter of a millimeter \cite{LCP}. However, the SM fields are confined to a thin (thickness not more than $10^{-17}$~cm \cite{brane_thick}) $D_3$-brane, which is a soliton solution of the underlying string theory on which the ends of open strings are confined. A crucial feature of this model is that gravity, which is a property of spacetime itself, is free to propagate anywhere in the bulk. On compactification, a Kaluza-Klein tower of closely spaced gravitons appear in our spacetime, a large number of which (controlled by $\sqrt{s}$) are producible\footnote{An alternative way of probing the ADD scenario is to consider virtual graviton exchange \cite{exdim_pheno} in SM processes where a coherent sum over closely spaced gravitons is involved, leading to deviations from SM predictions} in a collider process \cite{exdim_pheno} but are then invisibly lost in the higher dimensional bulk. To an observer on the brane, they would appear to be escaping unobserved with a large missing energy. This is a direct production of a three-body final state unlike the SUSY case where the decays of the heavy sleptons tend to generate more isotropic events.

%====================================================================%
\noindent
{\underline{\em Comparison of the two signals}}:~
Recall that our process is $e^+ e^- \to \l^+ \l^- \mise$ where $\l$ sums over both $e$ and $\mu$. Charged slepton ($\tilde e_{L,R}$ or $\tilde \mu_{L,R}$) pair production in an $e^+ e^-$ collider with both unpolarised and polarised beams has been explored earlier \cite{slep_pair}. Once produced, the sleptons decay into either a chargino-neutrino pair or into a neutralino-lepton pair. The partial decay widths are governed by both the mass and the composition of the charginos (neutralinos) as well as by the type ($L$ or $R$) of slepton. We select the channels yielding the final state of a same-flavour unlike-sign dilepton associated with a missing energy\footnote{In case $m_{\slep} > M_{\tilde{\chi}^\pm}$, there is also the chain $e^+ e^- \to  \slep^+_{L,R} \slep^-_{L,R} \to \tilde{\chi^+} \tilde{\chi^-} \nu_{\l} \bar{\nu_{\l}} \to \l^+ \l^- \neu \neu \nu_\l \nu_\l \bar{\nu_{\l}} \bar{\nu_{\l}}$. However, it makes a very small contribution, which we do take into account.}, namely
\begin{equation}
e^+  e^- \to  \slep^+_{L,R} \slep^-_{L,R} \to \l^+ \l^- \neu \neu .\\
\end{equation}
In our analysis, we do not adhere to any particular SUSY-breaking scenario and make no assumption related to any high scale physics other than adopting gauge coupling unification. Thus, whereas the slepton masses\footnote{Again, for simplicity, we take $m_{\slep_L} = m_{\slep_R} = m_{\slep}$ .} $m_{\tilde \ell}$ are free parameters in our analysis, the neutralino masses and couplings are completely specified by the respective $SU(2)$ and $U(1)$ gaugino masses $M_2$ and $M_1$, the Higgsino mass parameter $\mu$ and $\tan \beta$, which is the ratio \cite{Drees:2004jm} of the two Higgs vacuum expectation values arising in the MSSM.

The branching ratio for slepton decay into the lightest neutralino and the corresponding lepton depends on quite a few parameters: $\msl$, $\mu$, $\tan \beta$ as well as the gaugino mass parameters $M_1$ and $M_2$. Of these, the dependence on $\tan \beta$ is the least pronounced and therefore we shall henceforth use only one value of it, namely, $10$. Thus, only four parameters remain, namely $\msl$, $\mu$, $M_2$ and $M_1$. For a given slepton mass, the relevant branching fraction is then governed essentially by two factors: ($i$) the composition of the LSP and ($ii$) the energywise accessibility of slepton decay channels into the heavier neutralinos/charginos. The resulting dependence is still quite intricate and can be followed from ref~\cite{konar_slep}.

Turning to the ADD scenario, the production of a dielectron or dimuon pair with missing energy has been considered \cite{konar_add,mmg} earlier for probing its parameter space. The relevant reactions are
\begin{equation}
e^+ e^- \to  \l^+ \l^- G_n ,\\
\end{equation}
where $n$ is summed incoherently over the energywise accessible part of the tower of closely spaced gravitonic modes. Two parameters, determining the relevant cross-sections, are: ($i$) the number of extra dimensions $d$ and ($ii$) Planck's constant in the bulk or the so-called higher dimensional string scale $M_S$, expected to be in the TeV range.

The SM backgrounds to our signal (of a same-flavour, unlike-sign dilepton pair plus a substantial amount of missing energy) arise from all processes of the form 
\begin{equation}
e^+ e^- \to \l^+ \l^- \nu_i \bar{\nu_i}
\end{equation}
where $i$ can be any flavor. A significant portion of this background originates from the $\l \l Z$ final states, with the real $Z$ boson decaying into a neutrino pair carrying missing energy. These can be easily removed by imposing a suitable cut on the $\nu\bar{\nu}$ (missing) invariant mass. On the other hand, the background from $e^+ e^- \to W^+ W^-\to \l^+ \l^- \nu_\l \bar{\nu_\l}$ can be explicitly subtracted by reconstructing (with a two-fold ambiguity) events with the on-shell W-pair. Such a procedure is, of course, problematic if there is an accompanying photon or if one of the W's is off-shell. But then the use of appropriate longitudinally polarised beams would lead to a drastic reduction of this type of background. Though we perform our present analysis with unpolarised beams, we shall comment on the use of beam polarisation at the end.

%..................................................................
\begin{table}[t]
\begin{center}
$$
\begin{array}{|c|c|c|c|c|c|c|c|c|c|c|c|}
\hline
\multicolumn{6}{|c|}{\bf \sigma_{SUSY} [fb]}  &|&  \multicolumn{5}{|c|}{\bf \sigma_{ADD} [fb]}\\
\cline{1-6} \cline{8-12}
\multicolumn{2}{|c|}{\bf \tan \beta = 10}    &  \multicolumn{4}{|c|}{\bf \msl [GeV]}  &|&   \multicolumn{4}{|c|}{\bf M_S [TeV]}  & \\
\cline{1-6} \cline{8-11}
{\bf M_2, M_1 [GeV]} & {\bf \mu [GeV]} & \bf 155  & \bf 205  & \bf 225  & \bf 245     &|&    \bf .75  & \bf 1.0    & \bf 1.5  & \bf 2.0  & \bf d   \\ 
\cline{1-6} \cline{8-12}                
\bf 200, 100   &  \bf -400    &  427   &  164     &   59     &   7.8   &|&   1090  &  345      &  68    &   22 & \bf 2   \\
\cline{1-6} \cline{8-12}                 
\bf 300, 150   &  \bf -400    &  144   &  137     &   75     &   19    &|&   455   &  108      &  14    &   3.3 & \bf 3    \\
\cline{1-6} \cline{8-12}        
\bf 400, 200   &  \bf -150    &   92   &  40      &   13     &   0.6   &|&  202   &  36       &  3.2   &   0.6 & \bf 4  \\
\cline{1-6} \cline{8-12}
\bf 400, 200   &  \bf -100    &   79   &  32      &   6.9    &   0.3   &|&   97    &  13       &  0.8   &  0.1 & \bf 5\\
\hline
\end{array}
$$
$$
\begin{array}{|c|c|c|c|c|c|c|c|c|c|c|c|}
\hline
\multicolumn{6}{|c|}{\bf \sigma_{SUSY} [fb]}  &|&  \multicolumn{5}{|c|}{\bf \sigma_{ADD} [fb]}\\
\cline{1-6} \cline{8-12}
\multicolumn{2}{|c|}{\bf \tan \beta = 10}    &  \multicolumn{4}{|c|}{\bf \msl [GeV]}  &|&   \multicolumn{4}{|c|}{\bf M_S [TeV]}  &
\\
\cline{1-6} \cline{8-11}
{\bf M_2, M_1 [GeV]} & {\bf \mu [GeV]} & \bf 700  & \bf 800  & \bf 900  & \bf 1000     &|&    \bf 4.5  & \bf 5.0    & \bf 5.5  & \bf 6.0  & \bf d   \\
\cline{1-6} \cline{8-12}
\bf 200, 100   &  \bf -500    &   24   &   19     &   15     &    11
&|&   124  &  81      &  56    &   39 & \bf 2   \\
\cline{1-6} \cline{8-12}
\bf 400, 190   &  \bf -500    &   22   &   18     &   15     &   11    &|&   58   &  34      &  21    &   14 & \bf 3    \\
\cline{1-6} \cline{8-12}
\bf 600, 290   &  \bf -500    &   21   &   16     &   13     &   10    &|&   31   &  16       &  9.2   &   5.5 & \bf 4  \\
\cline{1-6} \cline{8-12}
\bf 800, 380   &  \bf -500    &   21   &   18     &   12     &    8    &|&   17    &  8.3       &  4.2   &  2.3 & \bf 5\\
\hline
\end{array}
$$
\caption[]{Cross-sections (in $fb$), after the imposition of the cuts described in the text, for the $e^+e^- \to \l^+ \l^- \mise$ signal in both ADD and SUSY scenarios at a linear collider with $\sqrt s$ = 500 GeV (upper panel) and  $\sqrt s$ = 3 TeV (lower panel). Parameter values are given in bold.}
\end{center}
\end{table}
%..................................................................

Let us consider an $e^+ e^-$ collider operating at a centre-of-mass energy\footnote{We here consider phase 1 of the ILC with CLIC taken up within brackets.} of 500~GeV (3~TeV). The kinematic cuts used in our analysis are as follows:
\begin{itemize}
\item Each of the final state charged leptons should be at least $10^0$ away from the beam pipe. This tames collinear singularities arising from $t$-channel photon exchange. At the same time, the elimination of any background effect from beamstrahlung is mostly ensured \cite{Godbole}.
\item Each charged lepton should have a transverse momentum $p_T^\l > 10$~GeV ($20$~GeV).
\item We demand a missing transverse momentum $p_T^{\rm miss} > 15$~GeV (25 GeV) since this also helps\footnote{{\sl cf.} $\S$ III of ref.~\cite{konar_add}} in reducing the two-photon background.
\item The tracks of the two unlike-sign leptons must be well-separated, with $\Delta R > 0.2$, where $\Delta R = \sqrt{\Delta\eta^2 + \Delta\phi^2}$ in terms of the differences of the lepton pseudorapidities $\Delta \eta$ and the azimuthal angles $\Delta \phi$.
\item The opening angle between the lepton tracks is required to be limited by $5^0 < \theta_{\l^+\l^-} < 175^0$. This ensures not only a sufficient missing energy, but also the elimination of possible cosmic ray backgrounds.
\item The missing invariant mass $M_{\rm miss}$ has to satisfy the inequality $|M_{\rm miss} - M_Z| > 10$~GeV in order to eliminate the background of two final state neutrinos arising from $Z$-decay. A further increase in the value of the lower cut in $M_{\rm miss}$ will reduce other SM backgrounds (such as those from off-shell $Z$ and $W^+ W^+$ production). On the other hand, too high a value of $M_{\rm miss}$ will reduce the signal. We have chosen a lower cut of 150 GeV (450 GeV) on $M_{\rm miss}$ which should reduce the SM background and at the same time yield a reasonable signal. 
\end{itemize}

The total cross-sections for our signal in the SUSY and ADD cases are presented in Table~1 (with sample choices of parameters) after applying the event selection criteria described above. We see that these can be comparable in magnitude for sizable parametric regions.
 For the SUSY case, we have checked that our selectron pair-production
cross sections match with what are plotted by Freitas et al. \cite{slep_pair} in
their Fig. 3(a) and moreover that the branching fractions for the decays
of charged sleptons into charged leptons plus LSP, used as multiplying
factors, agree with those of Choudhury et al. \cite{konar_slep}. A similar check for
the ADD case has been made with the results of \cite{konar_add}.
 The total contribution from SM backgrounds after applying the same event selection criteria comes out to be $36.4 \,fb$ ($72 \,fb$).  One can estimate the minimum value of the signal strength for which significance, defined by $S/\sqrt{B}$, where $S \, (B) =$ number of signal (background) events, takes the desired value of 3. Considering an integrated luminosity of $100$~fb$^{-1}$ ($1000$~fb$^{-1}$), the minimum signal strength needed for our analysis is found to be $\sigma_{S}=1.8 \, fb$ ($0.8 \, fb$).

%====================================================================%
\noindent
{\underline{\em Event shape variables}}:~
Though event shape analyses have so far been fruitful with many particle final states, even for two visible final state leptons, associated with a large missing energy, we expect to utilize the same to distinguish the signal of supersymmetry from the one originating from large extra dimensions of the ADD model. The underlying idea is the following. In the supersymmetric scenario a pair of rather heavy sleptons ($\slep^+ \slep^-$) are produced not far from the threshold in the center of mass energy of the collider. Each of these then subsequently decays into a lepton and a LSP. Because of the lack of any significant boost for each slepton, the daughter leptons lead to more isotropic events in the CM frame. The heavier the slepton, as compared to the LSP mass, the more isotropic is the event. In the ADD scenario, on the other hand,  the two leptons are produced in association with a single graviton; in fact, a significant part of the cross-section comes from the production of heavier graviton modes which are energywise accessible. Hence, the events here are more spiked. It should be noted further that the event shape would also carry some signature of the spin information of the graviton.

%..................................................................
\begin{figure}[t]
\centerline{
\epsfxsize= 8.0 cm\epsfysize=6.0cm
                     \epsfbox{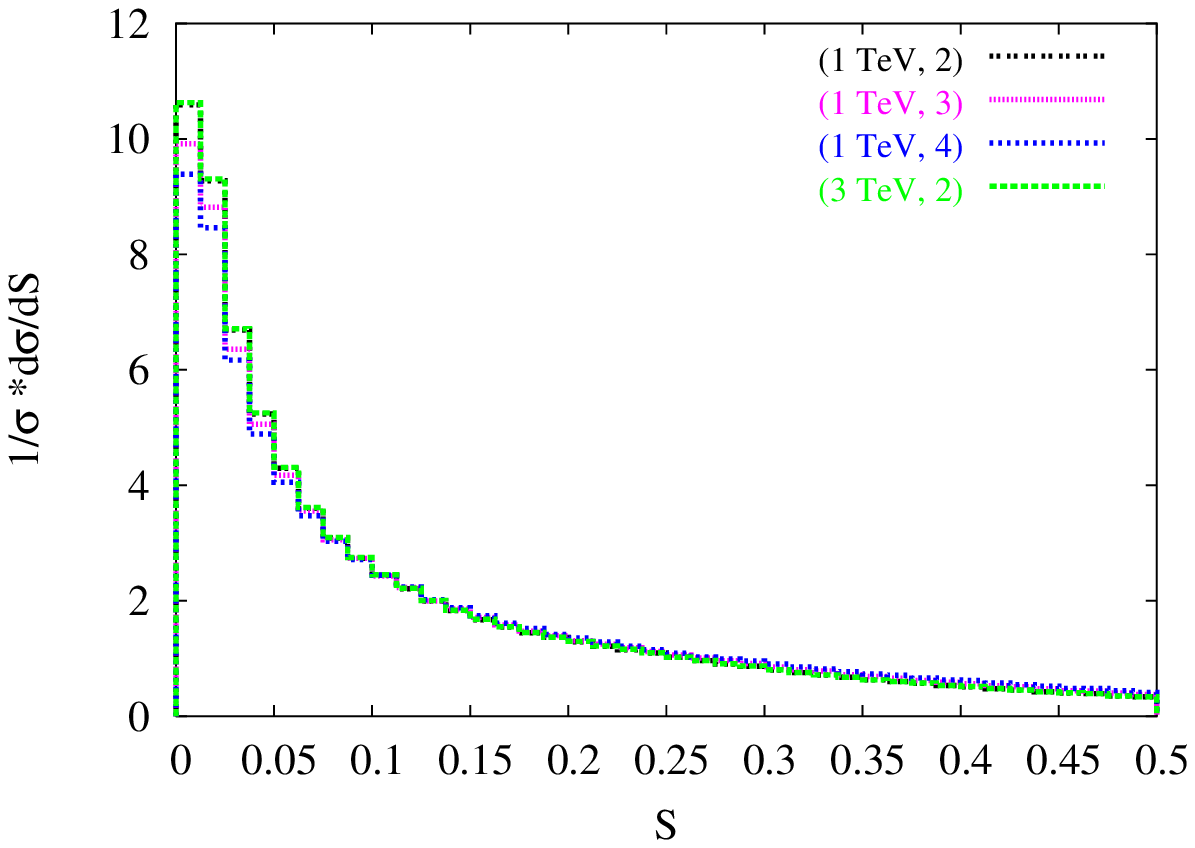}
        \hspace*{-.2cm}
\epsfxsize= 8.0 cm\epsfysize=6.0cm
                     \epsfbox{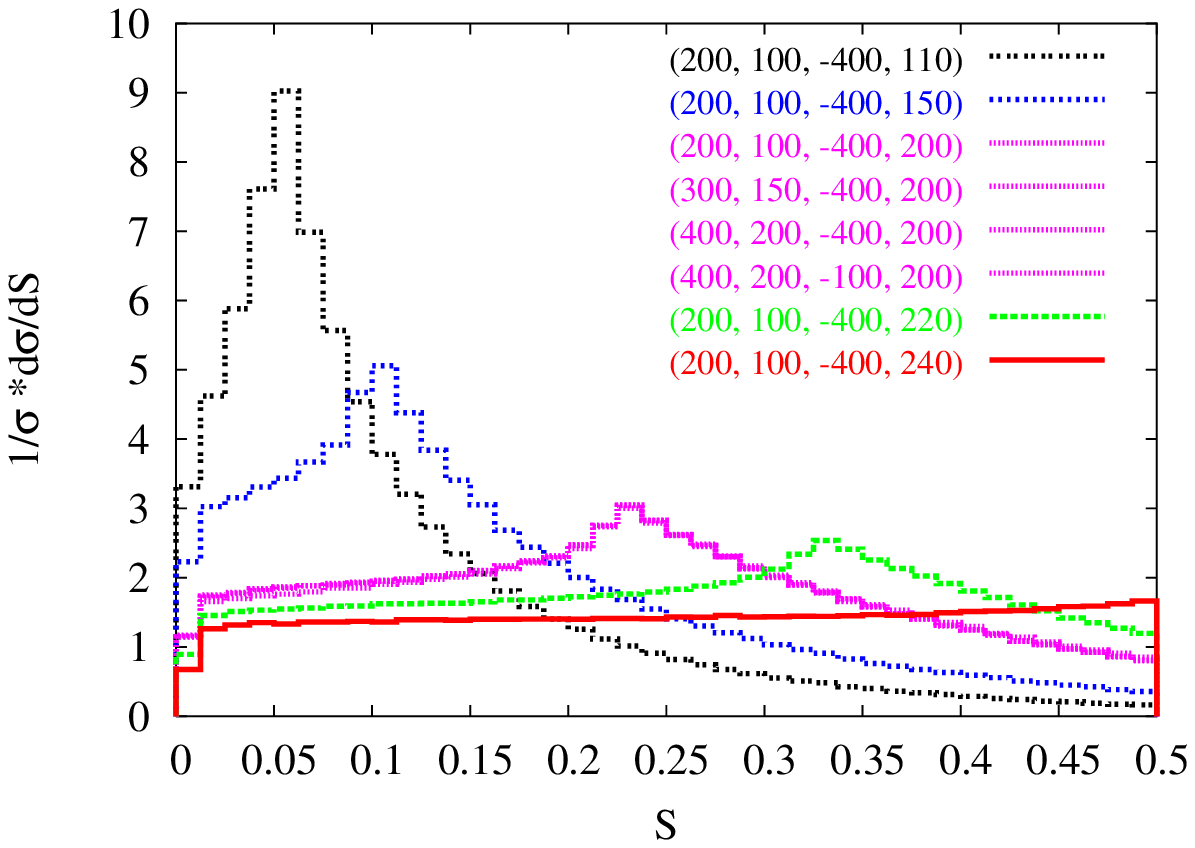}
}

\centerline{
\epsfxsize= 8.0 cm\epsfysize=6.0cm
                     \epsfbox{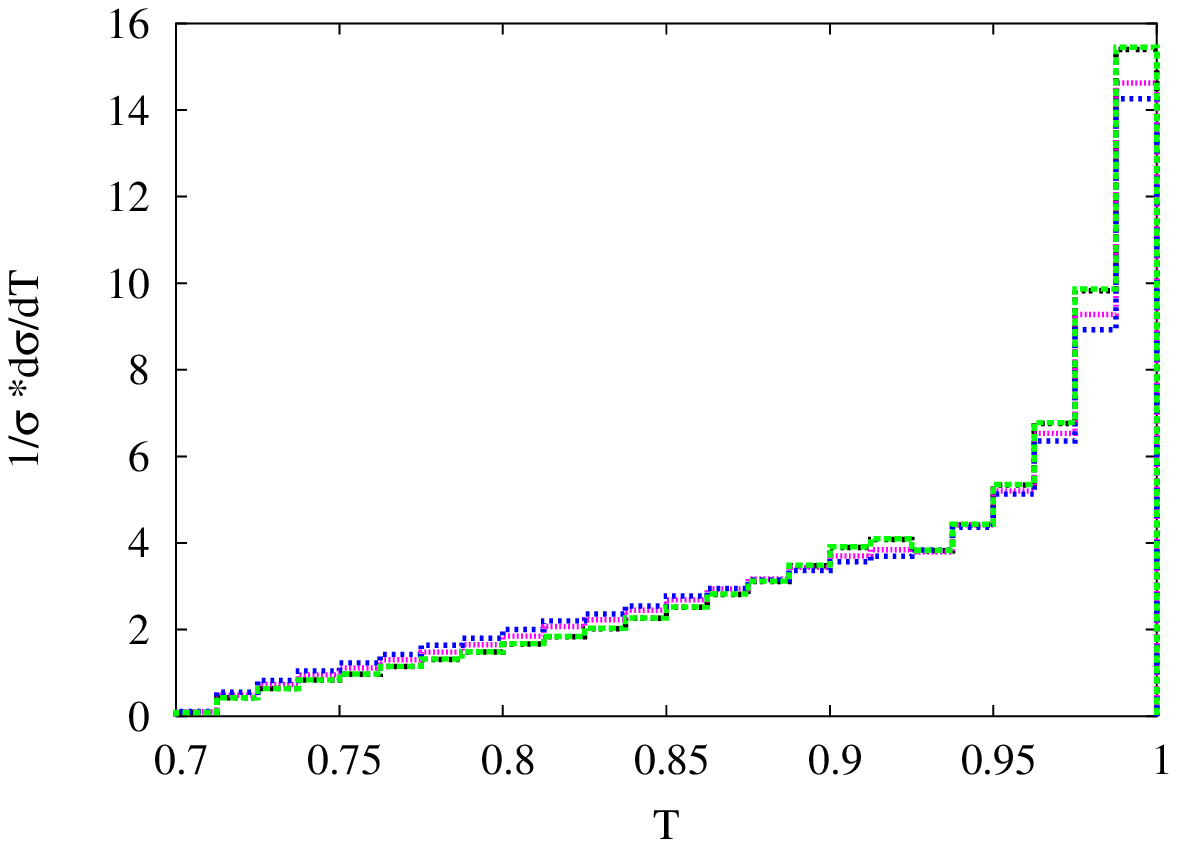}
        \hspace*{-.2cm}
\epsfxsize= 8.0 cm\epsfysize=6.0cm
                     \epsfbox{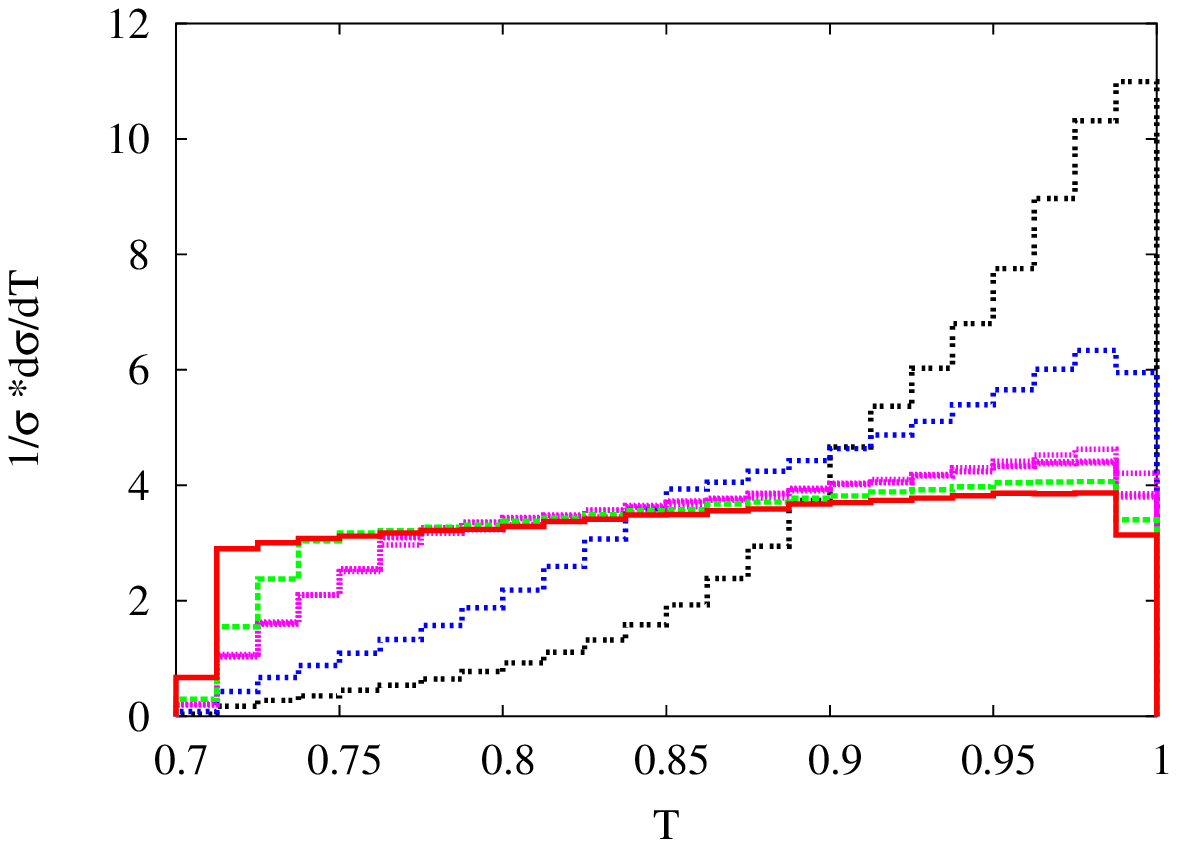}
}

\caption{\it Distributions of sphericity $S$ (upper panels) and  thrust $T$ (lower panels) for the $e^+ e^- \to  \l^+ \l^- \mise$ signal at $\sqrt{s}=500$ GeV both in the ADD (left panels) and SUSY (right panels) scenarios. The distributions are shown for different choices of parameters ($M_2$, $M_1$, $\mu$, $\msl$) in GeV for SUSY and ($M_S$, $d$) in the ADD case.}
\label{fig:mc}
\end{figure}
%......................................................................

Event shape characteristics can be specified by constructing variables like sphericity, thrust, oblateness \cite{collider_Barger} and circularity \cite{Barger}. It turns out to be sufficient for us to consider the first two. Criteria, made with such variables, have earlier proved to be promising, for instance, in reducing multiple jet backgrounds to the $t \bar{t}$ signal at the Tevatron \cite{Barger}. Sphericity is constructed from a normalised tensor,
\begin{equation}
S^{ij} = \frac{\sum_{a}p_a^i p_a^j}{\sum_{a}|\vec{p}_a|^2},
\end{equation}
where, $p_a^i$ is the $i$th component of the three-momentum $\vec{p}_a$ of the $a$th visible final state particle, with $i,j = 1,2,3$ and $a$ summed sum over all such particles. $S^{ij}$ has three eigenvalues $\lambda_{1,2,3}$ with $\lambda_1 \ge \lambda_2 \ge \lambda_3 \ge 0$ and  $\lambda_1 + \lambda_2 + \lambda_3 = 1$. Since only two of the $\lambda$'s are independent, the sphericity of an event can be defined as $S = \frac{3}{2} (\lambda_2 + \lambda_3)$. This is essentially a measure of the summed square of transverse momenta with respect to the event axis, two extreme cases being $S = 1$ for for an ideal spherical event and $S = 0$ for a linear event. On the other hand, thrust is defined by the quantity,
\begin{equation}
T = max \frac{\sum_a{|\vec{n}.\vec{p}_a|}}{\sum_a{|\vec{p}_a|}},
\end{equation}
where $|\vec{n}|$=1 and this vector $\vec{n}$ is the thrust axis for which maximum is attained. The allowed range is $\frac{1}{2} \le T \le 1$, where a spiked shape event has $T \sim 1$ and an isotropic event corresponds to $T \sim \frac{1}{2}$.

%====================================================================%
%..................................................................
\begin{figure}[t]
\centerline{
\epsfxsize= 8.0 cm\epsfysize=6.0cm
                     \epsfbox{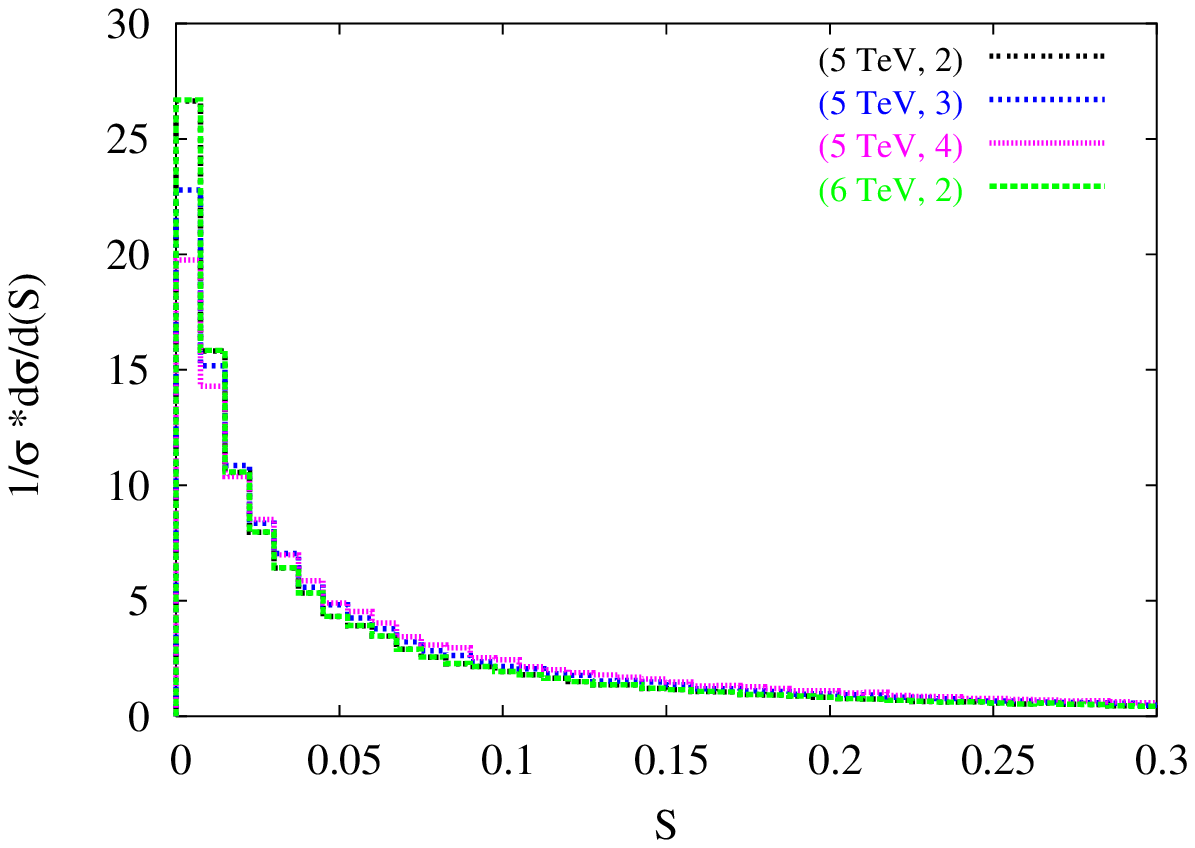}
        \hspace*{-.2cm}
\epsfxsize= 8.0 cm\epsfysize=6.0cm
                     \epsfbox{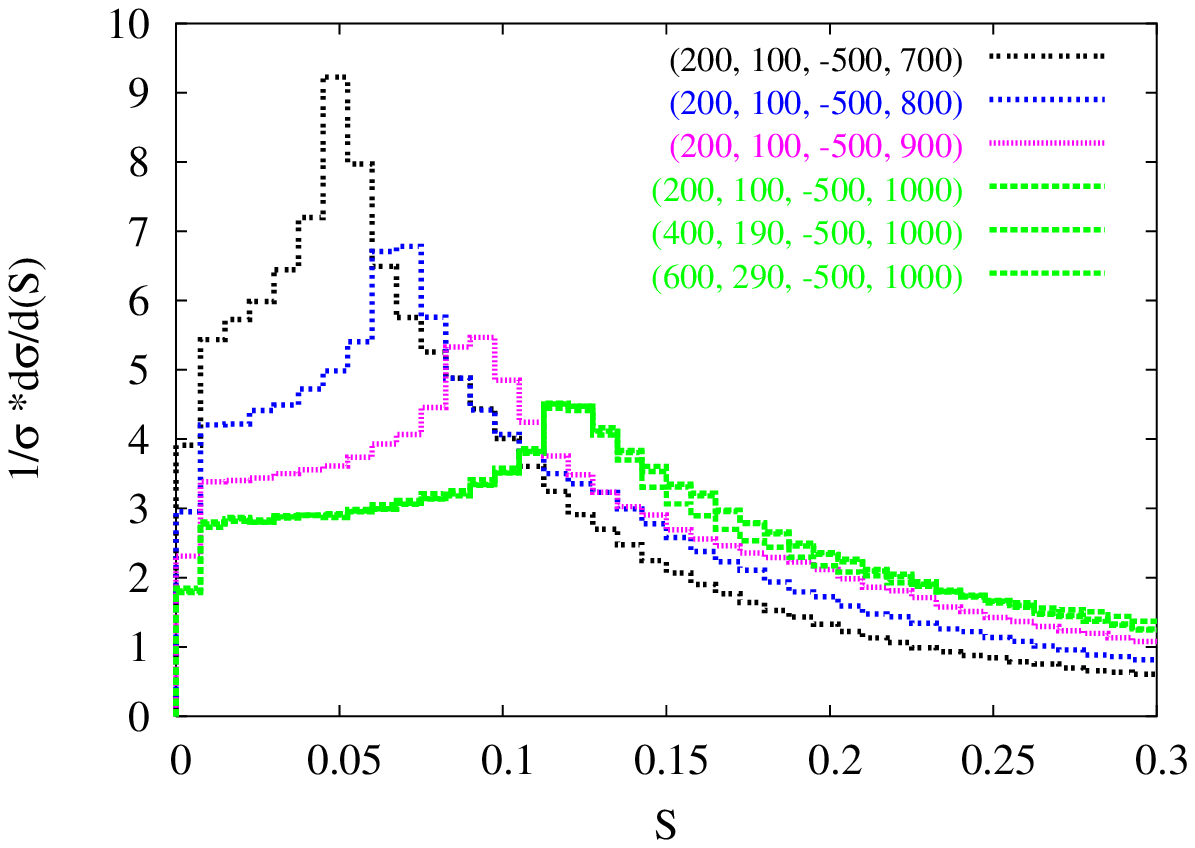}
}

\centerline{
\epsfxsize= 8.0 cm\epsfysize=6.0cm
                     \epsfbox{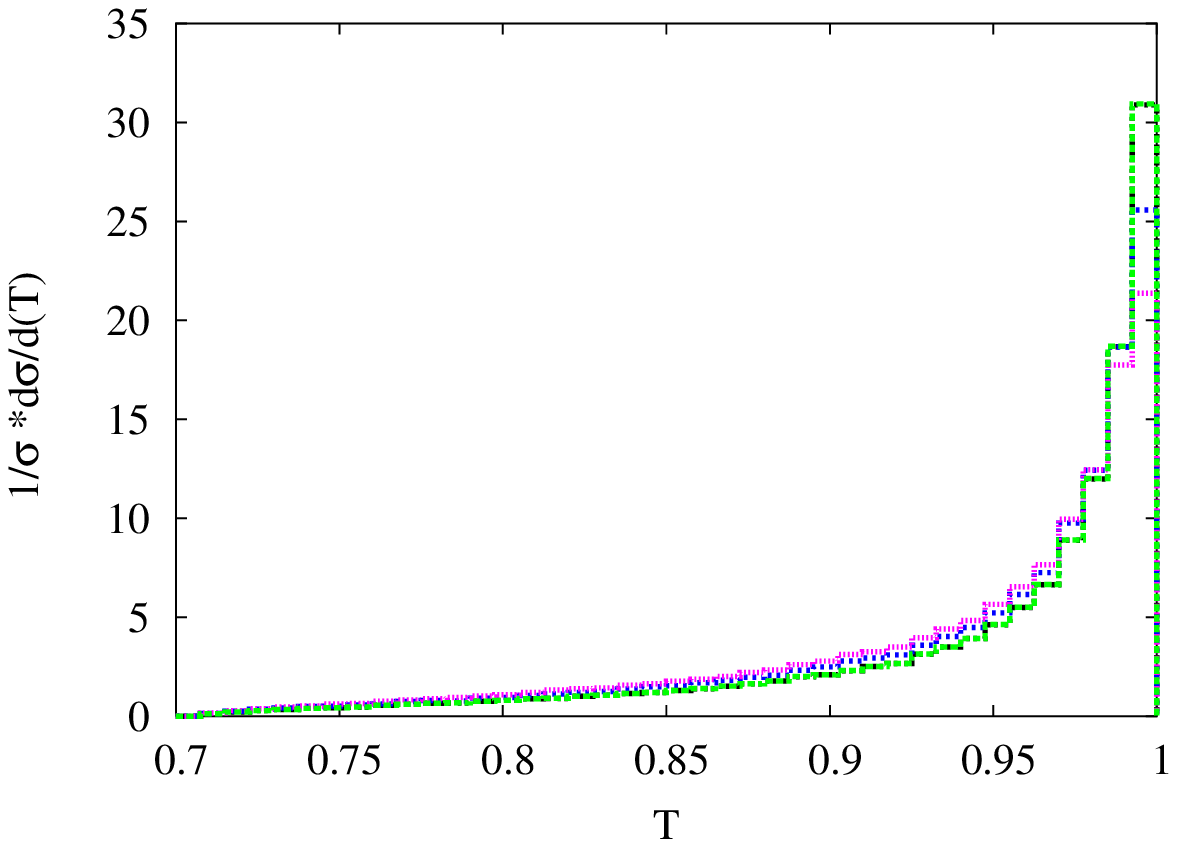}
        \hspace*{-.2cm}
\epsfxsize= 8.0 cm\epsfysize=6.0cm
                     \epsfbox{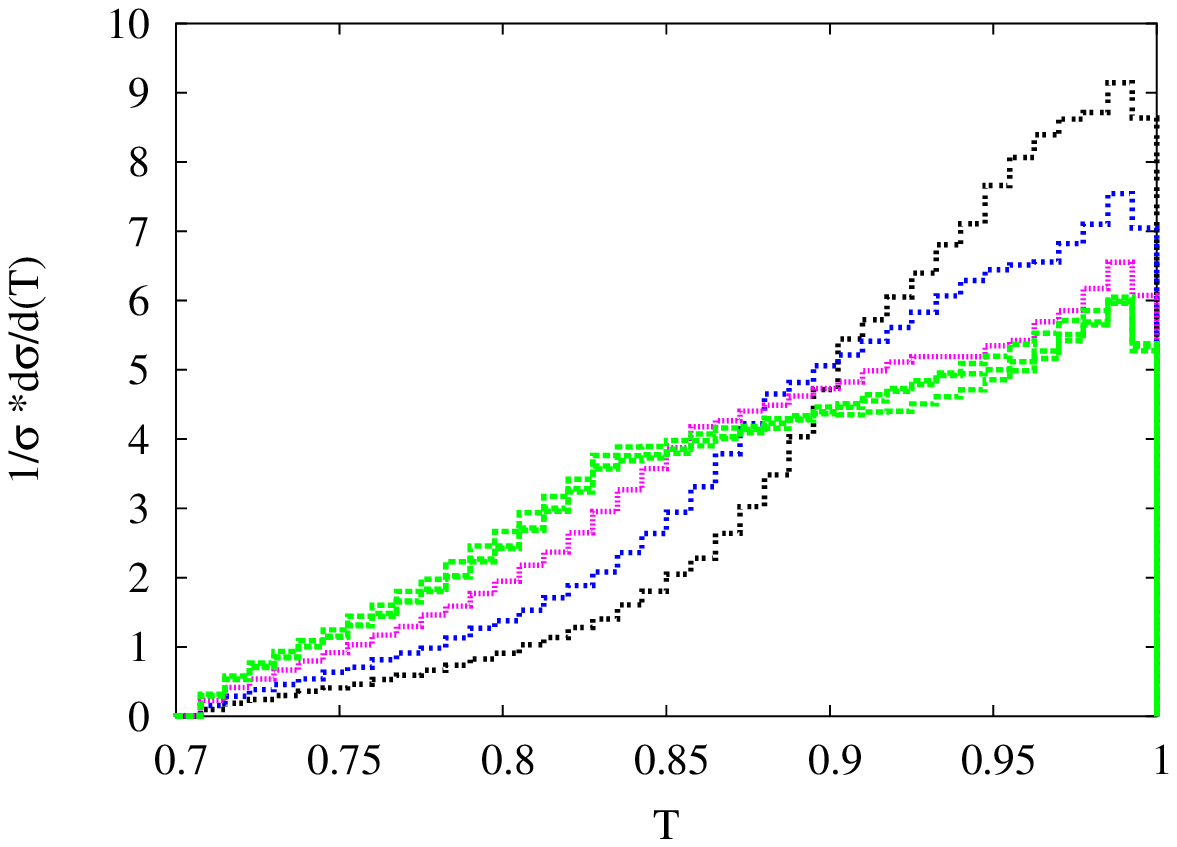}
}

\caption{\it Distributions of sphericity $S$ (upper panels) and  thrust $T$ (lower panels) for the $e^+ e^- \to  \l^+ \l^- \mise$ signal at $\sqrt{s}=3$ TeV both in the ADD (left panels) and SUSY (right panels) scenarios. The distributions are shown for different choices of parameters ($M_2$, $M_1$, $\mu$, $\msl$) in GeV for SUSY and ($M_S$, $d$) in the ADD case.}
\label{fig:clic}
\end{figure}
%......................................................................

%.....................................................................
\begin{figure}[t]
\centerline{
\epsfxsize=8.0 cm\epsfysize=7.0cm
                     \epsfbox{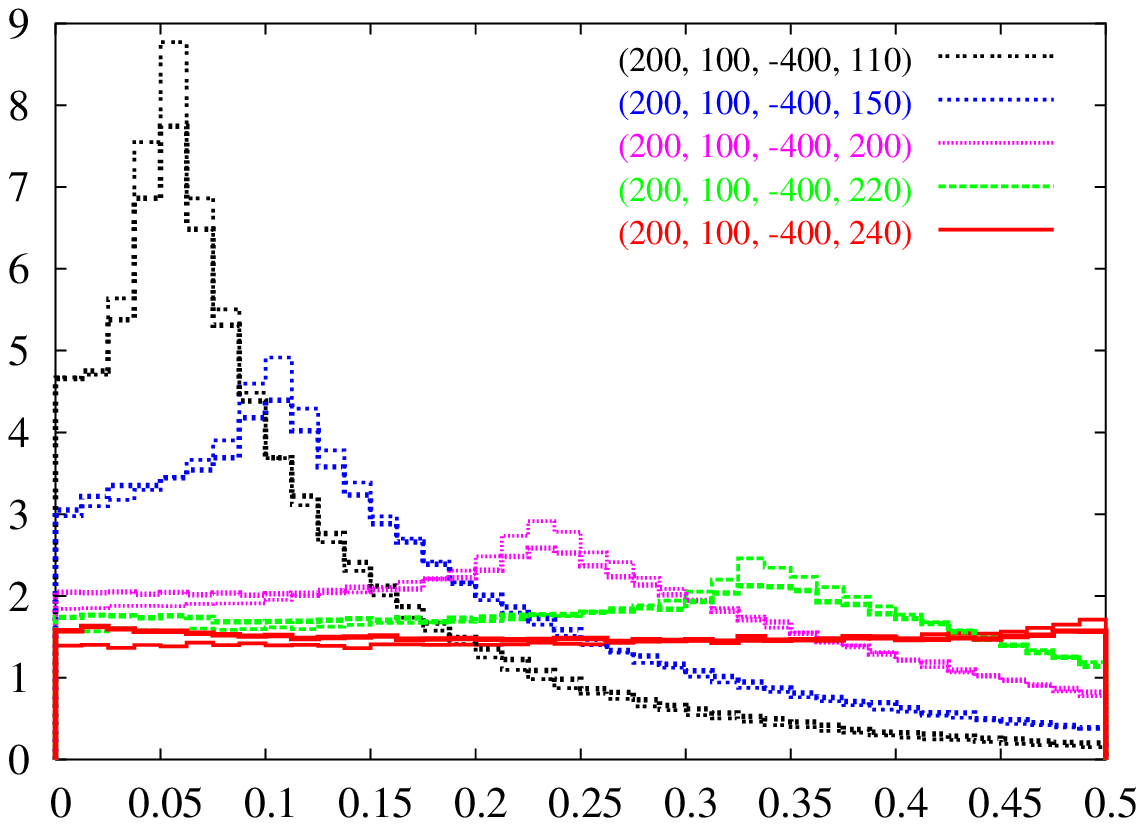}
        \hspace*{-.2cm}
\epsfxsize=8.0 cm\epsfysize=7.0cm
                     \epsfbox{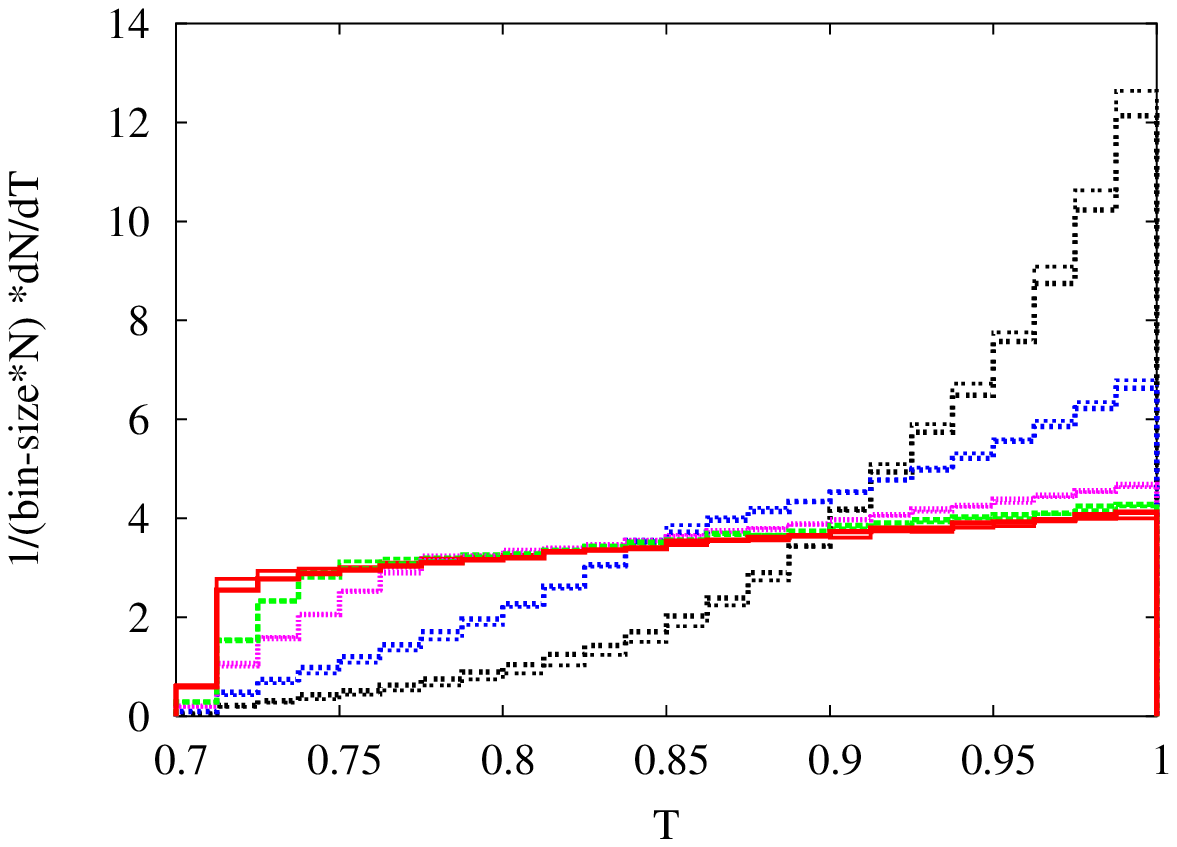}
}

\caption{\it Sphericity $S$ (left panel) and  thrust $T$ (right panel) distributions for the $e^+ e^- \to  \l^+ \l^- \mise$ signal at $\sqrt{s}=500$ GeV in the SUSY case reproduced in {\sc pythia} to include ISR and FSR effects (thin lines are without ISR and FSR effects). The different choices of parameters in the distributions are as in Fig.~1}
\label{fig:py}
\end{figure}
%.......................................................................

\noindent
{\underline{\em Results and Discussions}}:~
We now analyse the sphericity ($S$) and thrust ($T$) distributions for our signal computed with tree level diagrams. The computation has been performed with the {\sc comphep} \cite{chep} program package in the supersymmetric case and with the {\sc helas} subroutine \cite{helas} in the ADD case. In Figs.~1 and~2 we present these distributions for different choices of parameters. First, we note that, since there are only two visible particles here, the events are planer ($\lambda_3 = 0$) with the shape being circular rather than spherical for\footnote{With the ISR/FSR on, $S$ can go all the way to 1.} $S_{max}=\frac{3}{4}$. The sphericity distributions in the case of SUSY are seen to clearly depend on the slepton mass $m_{\slep}$, events become more and more circular for larger and larger values of $m_{\slep}$, showing a peaked structure, the peak location shifting towards the right as $m_{\slep}$ increases. On the other hand, sphericity distributions in the ADD case show a strong maximum at $S=0$, monotonically falling faster with $S$, being largely insensitive to the values of $M_S$ and $d$. We find in the SUSY case also that the sphericity distribution which is controlled\footnote{Therefore, the {\it value of the slepton mass $\msl$ can be extracted} from a measurement of the location of the sphericity peak.} by the slepton mass $\msl$, is insensitive to variations in $M_2$,\,$M_1$,\,$\mu$. Coming to thrust distributions, we see that they peak at  $T=1$ for the ADD case and again do not change much if $M_S$ and $d$ are varied. In the SUSY case, the thrust distributions are flatter except when $m_{\slep}$ is close to the present lower bound in which case they tend to resemble the ADD distributions. It is noteworthy that, even for $m_{\slep}$ as low as 110 GeV, the sphericity peak in the SUSY case is distinguishable from the maximum at $S=0$ for ADD. It is, however, true that the thrust distributions do not yield any additional advantage over the sphericity ones in so far as the basic discrimination is concerned. We have also looked at oblateness and circularity distributions which show similar features with no additional advantage. We have checked that similar features characterize the $\sqrt s$ = 3 TeV case.
%.....................................................................
\begin{figure}[t]
\centerline{
\epsfxsize=8.0 cm\epsfysize=7.0cm
                     \epsfbox{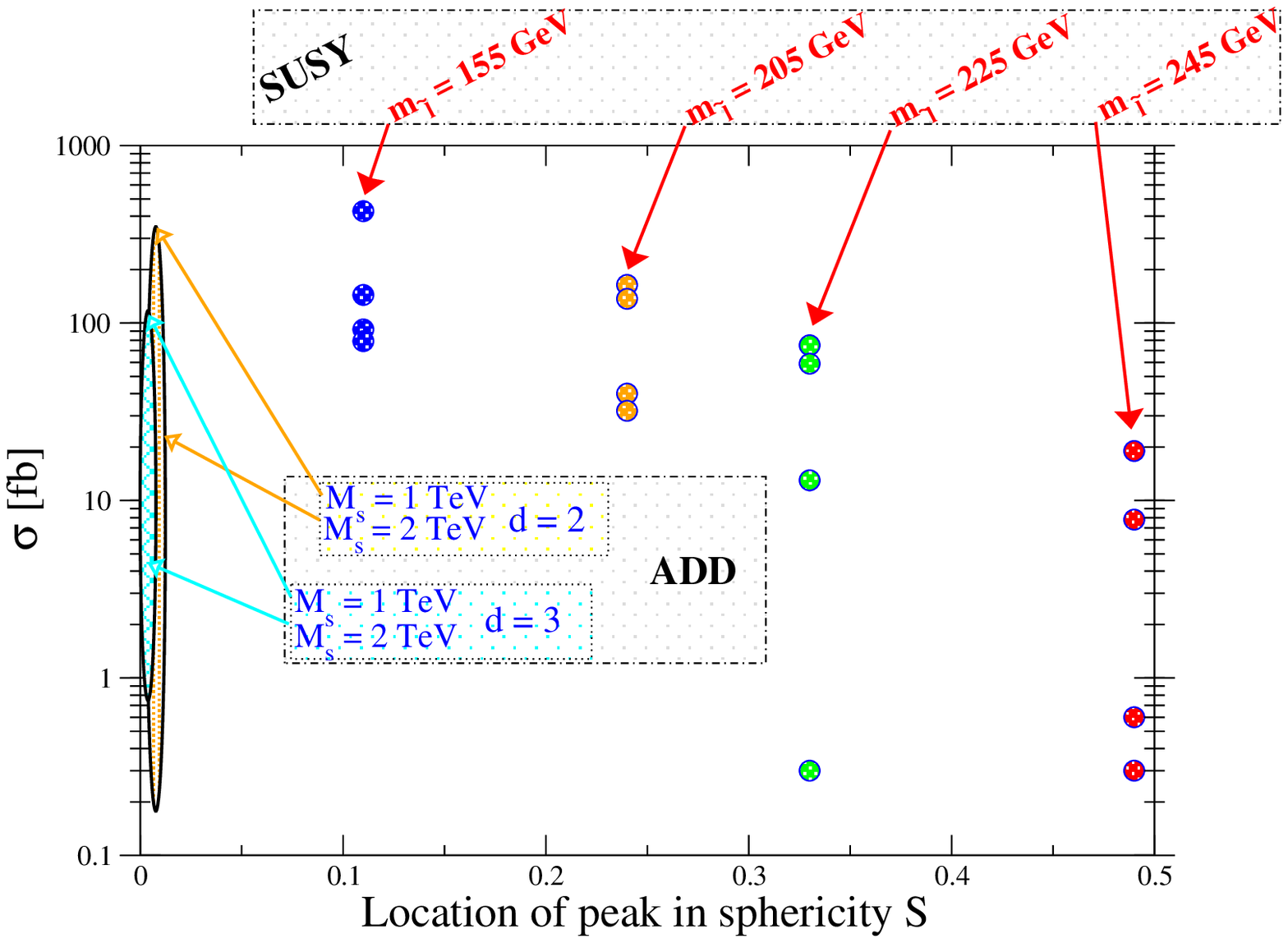}
}

\caption{\it Correlation between the total cross section and the location of the sphericity peak, for $\sqrt{s}=500$ GeV. Different choices of parameters are as in Table~1}
\label{fig:corr}
\end{figure}
%.......................................................................

One shortcoming of the above analysis may be that it has not included corrections due to the emission of collinear photons from the initial state (ISR) as well as the final state (FSR). Any sphericity distribution is regarded as vulnerable to changes caused by FSR because of the quadratic form of $S^{ij}$. Such an instability may be less pronounced for thrust which is linear in the concerned momenta. Though these effects are expected to be much smaller in leptonic processes than in hadronic ones because of the smallness of $\alpha_{EM}$ in comparison with $\alpha_{QCD}$, it is a concern that cannot be overlooked. We have, therefore, analysed these distributions for the SUSY case using the event generator {\sc pythia} \cite{pythia} which has the provision of including ISR and FSR contributions. The results, (with and without ISR and FSR) are shown in Fig.~3 for the $\sqrt s$ = 500 GeV case. Here the distribution of events, normalised by the total number of events times the bin size\footnote{We have chosen the same bin size for Fig.~1 and Fig.~3}, are considered for sphericity and thrust. We notice some small but visible changes in the sphericity distributions with and without ISR plus FSR -- but not enough to adversely affect the discrimination from the ADD case. As expected, all such changes are significantly less in the thrust distributions. 

A most important aspect of Fig.~3 is the following fact: unlike the shape of the lepton energy spectrum \cite{Battaglia:2005ma}, the {\sl locations} of the sphericity peaks for the SUSY case are unaffected by ISR and FSR effects. Moreover these locations are the same\footnote{The agreement between the plots of Fig.~1 and the thin lines of Fig.~3 is a cross-check on the consistency between the calculations with {\sc comphep} and with {\sc pythia}.} in Figs.~1 and~3. This location is therefore a robust discriminant between the SUSY and ADD cases since, for the latter, the peak is always at $S=0$. Fig.~4 shows a correlation plot for $\sqrt s$ = 500 GeV between the total cross-sections for the process $e^+ e^- \to  \l^+ \l^- \mise$ and the locations of the sphericity peaks shown as scatter points for some sample parametric choices. While the cross sections for the SUSY and ADD scenarios overlap, the sphericity peaks are very distinctly apart. The maximum for the ADD case is at $S=0$, while for SUSY it is at a nonzero value of $S$ which is an increasing function of $\msl$ moving towards the limit $\frac{1}{2}$ when $\msl$ approaches  $\frac{\sqrt{s}}{2} $. We have rescaled this whole analysis by taking $\sqrt{s} = 1$ TeV with higher values of the concerned parameters and have found very similar results. Thus even if SUSY or a model of extra dimensions \'a la ADD is beyond the reach of the first phase of the ILC and is only accessible to its second phase or to CLIC, our method of discrimination should work. 

As mentioned earlier, the use of appropriate longitudinally polarised beams would suppress the WW background significantly and one could have a reach for larger parameter spaces in both the SUSY and ADD scenarios. However, the distinctive features of sphericity and thrust distributions, as noted here, would be unaffected since their origin has nothing to do with beam polarisation. Let us also comment on beamsstrahlung \cite{Chen:1991wd} effects which we have ignored.  Because of our lower cut on $p_T^{miss}$, beamsstrahlung photons going down the beamlines will not affect our analysis. Nevertheless, beamsstrahlung will cause a degradation in the effective value of the CM energy  $\sqrt s$. This will quantitatively affect precision measurements, such as the extraction of $\msl$ from the  location of the sphericity peak, causing a systenmatic error which will need to be  taken into account. However, from the estimates made in \cite{Chen:1991wd}, we deem it unlikely that this degradation will change the qualitative difference in the sphericity distribution between the SUSY and ADD scenarios, namely a peak structure in the former and a strutureless monotonic falloff in the latter.

%====================================================================%
\noindent
{\underline{\em Summary and Conclusions}}:~
In the location of the maximum in the sphericity distribution of the process $e^+ e^- \to  \l^+ \l^- \mise\!\!\!$, we have discovered a sharp and robust discriminant between the SUSY with ADD scenarios which should be utilisable either at ILC or at CLIC wherever these scenarios become accessible. Longitudinally polarised beams with appropriate helicities should help further by suppressing the SM background. Similar considerations can be extended to the subprocess \cite{Han:1999ne} $q \bar{q} \to  \l^+ \l^- E \!\!\!\!/_T ~$, accessible at the LHC. However, a careful calculation of QCD corrections to it, with particular reference to ISR gluons, will be required first. 

%====================================================================%
\bigskip

\centerline{\em Acknowledgments}
{\footnotesize We thank S.Chakrabarti, M.Guchait, D.P.Roy, S.P.Trivedi and especially M.Drees, X.Tata for helpful discussions. PK acknowledges active discussions with participants in the Study Group on Extra Dimensions at LHC, held at the Harish-Chandra Research Institute.}

%====================================================================%

%====================================================================%

\begin{thebibliography}{99}
%....................................................................%
\def\pr#1,#2,#3 { {\em Phys.~Rev.}        ~{\bf #1},  #2 (#3) }
\def\prd#1,#2,#3{ {\em Phys.~Rev.}        ~{\bf D#1}, #2 (#3) }
\def\prl#1,#2,#3{ {\em Phys.~Rev.~Lett.}  ~{\bf #1},  #2 (#3) }
\def\plb#1,#2,#3{ {\em Phys.~Lett.}       ~{\bf B#1}, #2 (#3) }
\def\npb#1,#2,#3{ {\em Nucl.~Phys.}       ~{\bf B#1}, #2 (#3) }
\def\prp#1,#2,#3{ {\em Phys.~Rept.}       ~{\bf #1},  #2 (#3) }
\def\zpc#1,#2,#3{ {\em Z.~Phys.}          ~{\bf C#1}, #2 (#3) }
\def\epj#1,#2,#3{ {\em Eur.~Phys.~J.}     ~{\bf C#1}, #2 (#3) }
\def\mpl#1,#2,#3{ {\em Mod.~Phys.~Lett.}  ~{\bf A#1}, #2 (#3) }
\def\ijmp#1,#2,#3{{\em Int.~J.~Mod.~Phys.}~{\bf A#1}, #2 (#3) }
\def\ptp#1,#2,#3{ {\em Prog.~Theor.~Phys.}~{\bf #1},  #2 (#3) }
\def\jhep#1,#2,#3{{\em J.~High~Energy~Phys.}~{\bf #1}, #2 (#3) }
%....................................................................%


\bibitem{Drees:2004jm}
  M.~Drees, R.~Godbole and P.~Roy,
  {\sl Theory and phenomenology of sparticles} (World Scientific, Singapore 2004)
    %: An account of four-dimensional N=1 supersymmetry in high energy physics,''
%\href{http://www.slac.stanford.edu/spires/find/hep/www?irn=6240364}{SPIRES entry}
  and references therein.

\bibitem{Perez-Lorenzana:2005iv}
  J.L.~Hewett and J. March-Russell on p 1056 of 
{\sl Review of Particle Physics}, S. Eidelman {\sl et al.}, Phys.\ Lett.\ B {\bf 592}, 1 (2004);
%--
  A.~Perez-Lorenzana,
  %``An introduction to extra dimensions,''
  arXiv:hep-ph/0503177
  %%CITATION = HEP-PH 0503177;%%
  and references therein.

\bibitem{LC}
  K.~Ackermann {\it et al.},
  {\sl Extended joint ECFA/DESY study on physics and detector for a linear e+  e-
  collider, Proc. Summer Colloq. Amsterdam, Netherlands,  April 4,
  2003,}
  DESY-PROC-2004-01;
%\href{http://www.slac.stanford.edu/spires/find/hep/www?r=desy-proc-2004-01}{SPIRES entry}
%{\it Prepared for 4th ECFA / DESY Workshop on Physics and Detectors for a 90-GeV to 800-GeV Linear e+ e- Collider, Amsterdam, The Netherlands, 1-4 Apr 2003}
%--
  C.~P.~W.~Group: E.~Accomando {\it et al.},
  %``Physics at the CLIC multi-TeV linear collider,''
  arXiv:hep-ph/0412251;
  %%CITATION = HEP-PH 0412251;%%
%--
  K.~Desch, J.L.~Hewett, A.~Miyamoto, Y.~Okada, M.~Oreglia, G.~Weiglein and S.~Yamashita,
  %``The linear collider physics case: International response to the technology
  %independent questions posed by the International Technology Recommendation
  %Panel,''
  arXiv:hep-ph/0411159.
  %%CITATION = HEP-PH 0411159;%%


\bibitem{ADD}
  I.~Antoniadis,
  %``A Possible New Dimension At A Few Tev,''
  Phys.\ Lett.\ B {\bf 246}, 377 (1990);
  %%CITATION = PHLTA,B246,377;%%
%--
  N.~Arkani-Hamed, S.~Dimopoulos and G.~R.~Dvali,
  %``The hierarchy problem and new dimensions at a millimeter,''
  Phys.\ Lett.\ B {\bf 429}, 263 (1998);
%  [arXiv:hep-ph/9803315];
  %%CITATION = HEP-PH 9803315;%%
%--
  I.~Antoniadis, N.~Arkani-Hamed, S.~Dimopoulos and G.~R.~Dvali,
  %``New dimensions at a millimeter to a Fermi and superstrings at a TeV,''
  Phys.\ Lett.\ B {\bf 436}, 257 (1998);
%  [arXiv:hep-ph/9804398];
  %%CITATION = HEP-PH 9804398;%%
%--
  N.~Arkani-Hamed, S.~Dimopoulos and G.~R.~Dvali,
  %``Phenomenology, astrophysics and cosmology of theories with sub-millimeter
  %dimensions and TeV scale quantum gravity,''
  Phys.\ Rev.\ D {\bf 59}, 086004 (1999).
%  [arXiv:hep-ph/9807344].
  %%CITATION = HEP-PH 9807344;%%


\bibitem{Gopalakrishna:2001iv}
  S.~Gopalakrishna, M.~Perelstein and J.~D.~Wells,
  %``Extra dimensions vs. supersymmetric interpretation of missing energy
  %events at a linear collider,''
in {\it Proc. of the APS/DPF/DPB Summer Study on the Future of Particle Physics (Snowmass 2001),} ed. N.~Graf,
  eConf {\bf C010630}, P311 (2001).
%  [arXiv:hep-ph/0110339].
  %%CITATION = HEP-PH 0110339;%%

\bibitem{Asakawa:2002ij}
  E.~Asakawa, K.~Odagiri and Y.~Uehara,
  %``Measuring the spin of invisible massive graviton excitations at future
  %linear colliders,''
  JHEP {\bf 0301}, 062 (2003).
%  [arXiv:hep-ph/0204243].
  %%CITATION = HEP-PH 0204243;%%


\bibitem{Bhattacharyya:2005vm}
  G.~Bhattacharyya, P.~Dey, A.~Kundu and A.~Raychaudhuri,
  %``Probing universal extra dimension at the International Linear Collider,''
  arXiv:hep-ph/0502031.
  %%CITATION = HEP-PH 0502031;%%

\bibitem{Battaglia:2005ma}
  M.~Battaglia, A.~K.~Datta, A.~De Roeck, K.~Kong and K.~T.~Matchev,
  %``Contrasting supersymmetry and universal extra dimensions at colliders,''
  arXiv:hep-ph/0507284.
  %%CITATION = HEP-PH 0507284;%%


\bibitem{Appelquist:2000nn}
  T.~Appelquist, H.~C.~Cheng and B.~A.~Dobrescu,
  %``Bounds on universal extra dimensions,''
  Phys.\ Rev.\ D {\bf 64}, 035002 (2001).
%  [arXiv:hep-ph/0012100].
  %%CITATION = HEP-PH 0012100;%%

\bibitem{slep_pair} 
  J.~L.~Feng and M.~E.~Peskin,
  %``Selectron studies at e- e- and e+ e- colliders,''
  Phys.\ Rev.\ D {\bf 64}, 115002 (2001);
%  [arXiv:hep-ph/0105100];
  %%CITATION = HEP-PH 0105100;%%
%--
  A.~Freitas, A.~von Manteuffel and P.~M.~Zerwas,
  %``Slepton production at e+ e- and e- e- linear colliders,''
  Eur.\ Phys.\ J.\ C {\bf 34}, 487 (2004);
%  [arXiv:hep-ph/0310182];
  %%CITATION = HEP-PH 0310182;%%
%--
  C.~Blochinger, H.~Fraas, G.~Moortgat-Pick and W.~Porod,
  %``Selectron pair production at e- e- and e+ e- colliders with polarized
  %beams,''
  Eur.\ Phys.\ J.\ C {\bf 24}, 297 (2002).
%  [arXiv:hep-ph/0201282].
  %%CITATION = HEP-PH 0201282;%%

\bibitem{LCP}
  J.~C.~Long, H.~W.~Chan, A.~B.~Churnside, E.~A.~Gulbis, M.~C.~M.~Varney and J.~C.~Price,
  %``Upper limits to submillimeter-range forces from extra space-time
  %dimensions,''
  Nature {\bf 421}, 922 (2003).
  %%CITATION = NATUA,421,922;%%

\bibitem{brane_thick}
  R.~N.~Mohapatra, S.~Nussinov and A.~Perez-Lorenzana,
  %``Large extra dimensions and decaying KK recurrences,''
  Phys.\ Rev.\ D {\bf 68}, 116001 (2003).
%  [arXiv:hep-ph/0308051].
  %%CITATION = HEP-PH 0308051;%%

\bibitem{exdim_pheno}
  G.~F.~Giudice, R.~Rattazzi and J.~D.~Wells,
  %``Quantum gravity and extra dimensions at high-energy colliders,''
  Nucl.\ Phys.\ B {\bf 544}, 3 (1999);
%  [arXiv:hep-ph/9811291];
  %%CITATION = HEP-PH 9811291;%%
%--
  T.~Han, J.~D.~Lykken and R.~J.~Zhang,
  %``On Kaluza-Klein states from large extra dimensions,''
  Phys.\ Rev.\ D {\bf 59}, 105006 (1999).
%  [arXiv:hep-ph/9811350].
  %%CITATION = HEP-PH 9811350;%%


\bibitem{konar_slep}
  D.~Choudhury, A.~Datta, K.~Huitu, P.~Konar, S.~Moretti and B.~Mukhopadhyaya,
  %``Slepton production from gauge boson fusion,''
  Phys.\ Rev.\ D {\bf 68}, 075007 (2003).
%  [arXiv:hep-ph/0304192].
  %%CITATION = HEP-PH 0304192;%%

\bibitem{konar_add}
  S.~Datta, P.~Konar, B.~Mukhopadhyaya and S.~Raychaudhuri,
  %``Bhabha scattering with radiated gravitons at linear colliders,''
  Phys.\ Rev.\ D {\bf 68}, 095005 (2003).
%  [arXiv:hep-ph/0307117].
  %%CITATION = HEP-PH 0307117;%%

\bibitem{mmg}
  O.~J.~P.~Eboli, M.~B.~Magro, P.~Mathews and P.~G.~Mercadante,
  %``Direct signals for large extra dimensions in the production of fermion
  %pairs at linear colliders,''
  Phys.\ Rev.\ D {\bf 64}, 035005 (2001).
%  [arXiv:hep-ph/0103053].
  %%CITATION = HEP-PH 0103053;%%

\bibitem{Godbole}
  W.~Kozanecki {\it et al.},
  %``Summary of the ECFA linear collider working group on beamstrahlung and
  %photon-photon interactions,''
  SLAC-PUB-10058 (1992).
%\href{http://www.slac.stanford.edu/spires/find/hep/www?r=slac-pub-10058}{SPIRES entry}

\bibitem{collider_Barger}
  V.~D.~Barger and R.~J.~N.~Phillips,
  {\sl Collider Physics} (Addison-Wesley, USA 1987).
%\href{http://www.slac.stanford.edu/spires/find/hep/www?irn=1869256}{SPIRES entry}

\bibitem{Barger}
  V.~D.~Barger, J.~Ohnemus and R.~J.~N.~Phillips,
  %``Event shape criteria for single lepton top signals,''
  Phys.\ Rev.\ D {\bf 48}, 3953 (1993).
%  [arXiv:hep-ph/9308216].
  %%CITATION = HEP-PH 9308216;%%

\bibitem{chep}
  E.~Boos {\it et al.}  [CompHEP Collab.],
  %``CompHEP 4.4: Automatic computations from Lagrangians to events,''
  Nucl.\ Instrum.\ Meth.\ A {\bf 534}, 250 (2004).
%  [arXiv:hep-ph/0403113].
  %%CITATION = HEP-PH 0403113;%%

\bibitem{helas}
  H.~Murayama, I.~Watanabe and K.~Hagiwara,
  %``HELAS: HELicity amplitude subroutines for Feynman diagram evaluations,''
  KEK-91-11 (1991).
%\href{http://www.slac.stanford.edu/spires/find/hep/www?r=kek-91-11}{SPIRES entry}

\bibitem{pythia}
  T.~Sjostrand, L.~Lonnblad, S.~Mrenna and P.~Skands,
  %``PYTHIA 6.3: Physics and manual,''
  arXiv:hep-ph/0308153.
  %%CITATION = HEP-PH 0308153;%%

\bibitem{Chen:1991wd}
P.~Chen,
%``Differential luminosity under multi - photon beamstrahlung,''
Phys.\ Rev.\ D {\bf 46}, 1186 (1992).
%%CITATION = PHRVA,D46,1186;%%

\bibitem{Han:1999ne}
  T.~Han, D.~L.~Rainwater and D.~Zeppenfeld,
  %``Drell-Yan plus missing energy as a signal for extra dimensions,''
  Phys.\ Lett.\ B {\bf 463}, 93 (1999).
%  [arXiv:hep-ph/9905423].
  %%CITATION = HEP-PH 9905423;%%


\end{thebibliography}
\end{document}